\documentclass[12pt]{article}
\usepackage{amsmath,amssymb,amsfonts,amsthm}
\title{On the quantum information entropies and squeezing associated with the eigenstates of isotonic oscillator}

\author{A Ghasemi$^1$, M R Hooshmandasl$^1$ and M K Tavassoly$^2$
\\
\footnotesize{1- Department of Mathematical Sciences, Yazd University, Yazd, Iran} \\ \footnotesize{2- Atomic and Molecular Group, Faculty  of Physics,
Yazd University, Yazd, Iran}
\\ \footnotesize{E-mail: hooshmandasl@yazduni.ac.ir; mktavassoly@yazduni.ac.ir}}

\begin{document}
\maketitle

\begin{abstract}

In this paper we calculate the position and momentum space information entropies for the
quantum states associated with a particular physical system, i.e. the isotonic oscillator
Hamiltonian. We present our results for its ground states, as well as for its excited states. We
observe that the lower bound of the sum of the position and momentum entropies expressed by
the Beckner, Bialynicki-Birula and Mycielski (BBM) inequality is satisfied. Moreover, there
exist eigenstates that exhibit squeezing in the position information entropy. In fact, entropy
squeezing, which occurs in position, will be compensated for by an increase in momentum
entropy, such that the BBM inequality is guaranteed. To complete our study we investigate the
amplitude squeezing in $x$ and $p$-quadratures corresponding to the eigenstates of the isotonic
oscillator and show that amplitude squeezing, again in $x$, will be revealed as expected, while
the Heisenberg uncertainty relationship is also satisfied. Finally, our numerical calculations of
the entropy densities will be presented graphically.
\end{abstract}

 {\bf Pacs:} 03.65.-w, 42.50.Lc, 42.50.Dv

{\bf Keywords:} Isotonic oscillator; quantum information entropy; BBM inequality; squeezed state.

\section{Introduction}
 The probability densities of position and momentum of a single-particle system
 in one dimension are expressed as
 \begin{equation}\label{prob-dens}
    \rho(x)=|\psi(x)|^2,\hspace{1cm} \xi(p)=|{\phi}(p)|^2,
 \end{equation}
 where $\psi(x)$ is the solution of
 time-independent Schr$\ddot{o}$dinger equation and ${\phi}(p)$ is its Fourier transform
 in momentum space.
 Using the above expressions, {\it Shannon information entropy} of position and momentum space
 entropies are respectively defined as follows \cite{shannon}
 \begin{equation}\label{pos}
   S(\rho)=-\int \rho(x)\ln\rho(x)dx,
 \end{equation}
 \begin{equation}\label{mom}
   S(\xi)=-\int \xi(p) \ln \xi(p)dp.
 \end{equation}
 The latter definitions have been used frequently in the literature \cite{majernik, dehesa, Arkadiusz Orlowski, Manko}.
 These entropies can be helpful in particular purposes, for instance one may point to the
 reconstruction of the charge and momentum densities of atomic and molecular systems \cite{galin, angul} by means of maximum entropy procedure.
 It is believed that the above entropies for one-dimensional systems lead to a stronger version of the Heisenberg uncertainty relation written
 as (see  p. 28 of  \cite{dodonov})
\begin{equation}\label{bbm}
   S(\rho)+ S(\xi)\geq \ln(e \pi),
\end{equation}
  which is a consequence of a well-known inequality in Fourier analysis, first conjectured by
  Everett in \cite{evert}, Hirschman  in \cite{dsa} and then proved by
  Bialynicki-Birula, Mycielski \cite{hash} and Beckner (BBM) in \cite{dsaa}.
  The inequality (\ref{bbm}) indicates that the  sum of entropies bounded from below by the value 2.1447...
  for one dimensional systems.
  This lower bound is obtained from the Gaussian wave function of the ground state of harmonic oscillator.
  It ought to be mentioned that the relation (\ref{bbm}) can be expressed for any two conjugate operators.
  For instance, recently we have examined the entropy uncertainty relation for the phase-number operators of
  nonlinear coherent states corresponding to solvable quantum systems with discrete spectra \cite{tavas}.

 The analytical form of position and momentum space entropies in (\ref{pos}) and (\ref{mom}) have been found
 for a few particular solvable quantum mechanical systems.
 For instance, the momentum and position entropies associated to ground state of harmonic oscillator were
 exactly calculated, where it is shown that the BBM inequality is saturated: $S(\rho) = 1.07236 = S(\xi)$
 (recall as a well-known fact that the vacuum state and the canonical coherent state also minimize the Heisenberg uncertainty
 relation). While for exited states of the harmonic oscillator the related entropies may be calculated only approximately \cite{majernik}.
 As second example, one may refer to position and momentum space entropies associated to
 P\"{o}shl-Teller potentials. These quantities for the ground
 state of this system are exactly evaluated,
 but even for the first excited states the results have been again given numerically
 \cite{poshteller}.
 Another physical system whose quantum information entropies were
 previously were considered and discussed is the Morse potential, for which all of the related results have been given numerically \cite{morse}.
 In all of the above mentioned systems the lower bound which predicted by the BBM inequality is guaranteed.
 On the other side, in addition to the above physical systems, Shannon information entropies for some classical orthogonal polynomials
 such as Hermite, Laguerre and Gegenbauer polynomials have also been introduced recently \cite{dehesa, coppele}.
 Recently, the evolution of entropy squeezing in quadratures of a single-mode field,  in the Jaynes-Cummings
 model, in the presence of nonlinear effect studied and discussed in
 \cite{Chaos}. Recently, the entropy squeezing is investigated using Shannon information
entropy for the solutions of the Hamiltonian describing the
interaction between a single two-level atom and two
electromagnetic fields in the framework of a modified
Jaynes-Cummings model \cite{IJQI}.

 The present contribution deals with the same studies, i.e., investigates the position and momentum entropies
 corresponding to a special quantum system known as  {\it isotonic oscillator}  \cite{Landau},
 through which we will find that the BBM inequality is satisfied.
 But, surprisingly, unlike the previously discussed cases in the literature \cite{majernik,  poshteller, morse},
 as we will observe in the continuation of
 the paper, the {\it "entropy squeezing"} occurs in
 position for some special eigenstates of the isotonic oscillator, i.e., it becomes less than that of the
 harmonic oscillator.
 Clearly, enough amount of increase in the momentum entropy,
 corresponding to the eigenstates whose entropy squeezing  in $x$ has been occurred,
 compensates the reduction of entropy in position, such that the BBM inequality  (\ref{bbm}) is not violated.
 Recall that in neither of the previously discussed physical systems in the literature, like
 harmonic oscillator, P\"{o}shl-Teller and Morse potentials entropy squeezing is not reported.
 Meanwhile, our considered system is a real physical potential, which possess entropy squeezing in position space.

 It was established in \cite{Arkadiusz Orlowski} that for all quantum states
having a squeezing property in one of the quadratures
(double squeezing is a very rare phenomenon [20, 21]), the
corresponding entropy is also squeezed. Hence, apart from
the above mentioned results on entropy squeezing emphasized
in this paper, in view of the connection between the BBM inequality and the Heisenberg uncertainty relationship, we
have calculated the variances of the two quadratures of
the field associated with the isotonic oscillator. In light of
our further numerical calculations, which are presented later
in the paper, we may say that the ground state of the
isotonic oscillator (as a real physical potential) can reveal
a quadrature squeezing feature; i.e. its spatial variance falls
below that of the vacuum. Therefore, our results confirm that
whenever squeezing in one of the quadratures is detected,
the corresponding information entropy is also squeezed.
Nevertheless, as expected there are states that possess entropic
squeezing in position, while quadrature squeezing is not seen.
This is also in agreement with the statement reported in \cite{{Arkadiusz Orlowski}}.
 In view of the increasing interest in the non-classical state
in quantum optics and related fields of research, we wish to
establish the non-classical features of some of the eigenstates
of the isotonic oscillator that have not been pointed out up
to now. In addition, we have found a set of (entropic or/and
quadrature) squeezed states corresponding to a well known
physical potential.
The paper is organized as follows. In the next section we
will address the entropies of the eigenstates of the isotonic
oscillator Hamiltonian. In section 3 we present the numerical
results and make some comments on the isotonic oscillator
system. Finally, in section 4 we outline the conclusion.

 \section{Entropies of the eigenfunctions  of isotonic oscillator}

 There exist a few potentials corresponding to physical systems
whose exact solutions are known. Among these solvable
systems, the isotonic oscillator is an interesting model with
Hamiltonian
 \begin{equation}\label{E1}
    H=-\frac{d^2}{dx^2}+x^2+\frac{A}{x^2},\hspace{1cm} A\geq0,
 \end{equation}
 acting in Hilbert space $L_2(0, \infty)$. The  Hamiltonian in (\ref{E1}) admits exact solutions as follows \cite{Landau, Hall},
 \begin{eqnarray}\label{E2}
   \psi_m^\gamma(x) &=&(-1)^m\sqrt{\frac{2(\gamma)_m}
   {m!\Gamma(\gamma)}}x^{\gamma-\frac{1}{2}}e^{-\frac{1}{2}x^2} {}_1F_1(-m;\gamma;x^2)\nonumber\\
   &=&\sum_{k=0}^m(-1)^m\sqrt{\frac{2(\gamma)_m}{m!\Gamma(\gamma)}}\frac{(-m)_k}{(\gamma)_kk!}x^
   {2k+\gamma-\frac{1}{2}}e^{-\frac{1}{2}x^2},
 \end{eqnarray}
  with the Dirichlet boundary condition $\psi_m^\gamma(0)=0$, where $\hspace{1mm} \gamma=1+\frac{1}{2}\sqrt{1+4A}\hspace{1mm}$
  and the eigenvalues are exactly obtained as \hspace{1mm}$e_m=2(2m+\gamma)$.
  Also notice that ${}_1F_1(-m;\gamma;x^2)$ denotes the Kummer confluent hypergeometric function,
  $\Gamma(\gamma)$ denotes the Gamma function and $\hspace{1mm}(\gamma)_m \doteq \gamma(\gamma+1)(\gamma+2)\cdots(\gamma+m-1)$ is the
  Pochhammer symbol with $(\gamma)_0=1$. The condition $A\geq0$ stated in (\ref{E1})
  is equivalent to $\gamma \geq 3/2$ in (\ref{E2}).
  It is obviously seen that adding the term $A/x^2$ to the harmonic oscillator
  Hamiltonian does not change the equidistant aspect of the spectrum of linear harmonic oscillator.
  A word seems to be necessary about the completeness of the eigenstates in (\ref{E2}) \cite{Hall}.
  It is proved by Hall {\it et al} that these  eigenfunctions satisfy the orthonormality condition
  in $L_2(0, \infty)$, that is $\int _0^\infty \psi_m^\gamma(x) \psi_n^\gamma(x)=\delta_{m, n}$
  where $m,n=0, 1, 2, 3, \cdots $.
  So the set of  $\left\{\psi_m^\gamma(x)\right\}_{m=0}^\infty$-functions is a
  complete orthonormal basis for the Hilbert space $L_2(0, \infty)$.

  The Fourier transform of $\psi_m^\gamma(x)$ in (\ref{E2})
  is denoted by us as $\phi_m^\gamma(p)$ and it is given by
   \begin{eqnarray}\label{tran}
   \phi_m^\gamma(p) &=&\frac{1}{\sqrt{\pi}}\sum_{k=0}^mC_k(\gamma,m)\left[\Gamma(\frac{1}{4}+k+\frac{\gamma}{2})
   {}_1F_1(\frac{1}{4}+k+\frac{\gamma}{2};\frac{1}{2};\frac{-p^2}{2})\right.\nonumber\\
   &+&\left.i\sqrt{2}p\Gamma(\frac{3}{4}+k+\frac{\gamma}{2}){}_1F_1(\frac{3}{4}+k+\frac{\gamma}{2};\frac{3}{2};\frac{-p^2}{2})\right]
 \end{eqnarray}
   where we set
   $$C_k(\gamma,m) \equiv \frac{(-1)^m2^{-\frac{3}{4}+k+\frac{\gamma}{2}}(-m)_k}{k!(\gamma)_k}\sqrt{\frac{(\gamma)_m}{\pi
   m!\Gamma(\gamma)}}.$$
   The term $\lambda^{-1/2}$ is a normalization factor may be determined for
   fixed $m, \gamma$,  i.e.,
   $\int_{-\infty}^{\infty}|\phi_m^\gamma(p)|^2 \,dp =1$.
   Inserting the expressions obtained in (\ref{E2}) and (\ref{tran}) in the following integrals
 \begin{equation}\label{x entropy}
 S_m^\gamma(\rho)=-\int_{0}^\infty|{\psi}_m^\gamma(x)|^2 \ln|{\psi}_m^\gamma(x)|^2dx,\hspace{3.5cm}
 \end{equation}
 \begin{equation}\label{p entropy}
 S_m^\gamma(\xi)=-\int_{-\infty} ^\infty|\phi_m^\gamma(p)|^2 \ln|\phi_m^\gamma(p)
 |^2dp, \hspace{1cm} m=0,1,2,\cdots
 \end{equation}
 yield the position and momentum information entropies \cite{majernik, dehesa, Arkadiusz Orlowski, Manko}.
 Note that while in the integration procedure in (\ref{x entropy}) the limits are from $0$ to $\infty$
 (in consistence with
 the corresponding Hilbert space $L_2(0, \infty)$), the limits of integration in (\ref{p entropy})
 are from $-\infty$ to $+\infty$. The latter one is
 due to the fact that the particle confined in the well-potential can travel back and forth (positive and negative).
 Seemingly, solving the above integrals in closed form to obtain the information entropies for arbitrary values of
 $\gamma$ and $m$ is very hard if not impossible, so we confine ourselves to some particular values of
 $\gamma$ and $m$ in the continuation of the paper.

\section{Numerical Results}

  We will now study the entropies of position and momentum
for the eigenstates of the isotonic oscillator Hamiltonian. In
general, the analytical solutions of the entropies defined in (\ref{x entropy}) and (\ref{p entropy})
   are not possible for arbitrary  values of $\gamma$ and $m$.
   Instead, one can work distinctly with for instance $\gamma =3/2, 5/2, 7/2, \cdots$,
   also fixed values of $m$ should be taken into account.
   The numerical results for the entropies in (\ref{x entropy}) and (\ref{p entropy}) and their sum, for three
   different values of $\gamma$ were expressed in Table 1, where four different values of $m$ are considered for each $\gamma$.
   From the Table 1 it is clear that for any choice of $\gamma$, the entropies and their sum increase
   with  increasing the values of $m$.
   Interestingly, for any fixed $\gamma$, there exists some
   values of $m$ (ground and first excited states) for which the information entropies in position
   become less than $1.07236$, i.e., the ground state's value of harmonic oscillator. In other
   words the entropy squeezing occurs in $x$.  Altogether,
   it is observed that the BBM inequality for arbitrary values  of $m$ and $\gamma$
   are satisfied in all cases for the states displayed in Table 1.
   Indeed, an increase in the momentum entropy,
   relative to that of the ground state of harmonic oscillator, is observed for any $\gamma$ (with $m=0, 1$),
   which is enough for compensating the reduction in position
   entropy.
   Consequently, our results confirm the stated result in \cite{Gonzaler}, i.e.,
   "there are no physical states which violate the BBM inequality".

   Nevertheless, due to the relationship that exists between
the BMM inequaliy and the Heisenberg uncertainty relation
we have been motivated to evaluate the variances of the
quadratures of the field defined as
   \begin{equation}
      (\Delta z)^2 = \langle z^2\rangle - \langle z \rangle^2,
      \qquad z=x  \quad {{or}} \quad p.
   \end{equation}
   Noticing that $[x, p]=i$, the Heisenberg uncertainty reads as $(\Delta x)^2 (\Delta p)^2 \leq 0.25$. Therefore,
   a state is squeezed if $(\Delta x)^2 < 0.5$ or $(\Delta p)^2 < 0.5$.
   In the squeezing phenomenon, the reduction in the uncertainties of one of the quadrature below the value of vacuum occurs
   with the cost of an increase in the conjugate quadrature.
   We have used (6) and (7)
   respectively for the calculation of squeezing in $x$ and $p$.
   Our displayed results in Table 2 on noise fluctuations show that for the ground
   states
   (which the entropy squeezing in position is occurred) squeezing in $x$-quadratures of the field is visible.
   Our further calculations show that for $\psi_{m=0}^\gamma (x)$ one has $(\Delta x)^2= 0.2427,$ $0.2441,
   0.2450$, and $(\Delta p)^2= 1.0714, 1.0555, 1.0455$
   respectively for $\gamma= 9/2,$ $ 11/2, 13/2$. Henceforth, the multiplication of $(\Delta x)^2(\Delta p)^2$ yields $\simeq0.25$.
   Continuing these computations, we observed that the squeezing effect in $x$ ($p$) will tend to the
   finite value $0.25$ ($1$) with increasing $\gamma$.
   Meanwhile, it ought to be mentioned that these states are evidently rather different from the
   well-known "squeezed states" in quantum optics, have been obtained from the action of the squeezing operator
   on the vacuum of the field.
   Henceforth, we may firstly pointed out that all of the "ground states" (with arbitrary $\gamma$) of the "isotonic oscillator",
   as a physical potential, are indeed  "squeezed states".
   Thus, since all of the ground states of the isotonic oscillator possess
   this property, independent of the values of $\gamma$, i.e., they
   can reasonably be considered as a class of (non-classical) squeezed states which are directly
   related to a well-known physical system.
   In addition, some special sets of eigenstates of isotonic oscillator with $m=0$ may be called the "ideal squeezed
   coherent states", since they saturate Heisenberg inequality (where we have used the definition of "ideal squeezed
   coherent states" as introduced by Scully and Zubairy in \cite{Scully and Zubairy}).
   At this stage, it is worth mentioning that it is recently proved that
   the quadrature squeezing exhibition is always accompanied by a corresponding entropy
   reduction below the vacuum entropy  level $\simeq 1.07236$ \cite{Arkadiusz Orlowski}.
   This feature was not be observed in the previously considered
   physical systems, like harmonic oscillator, P\"{o}shl-Teller and Morse potentials.


  To this end, instead of the evaluation of quantum information entropies,
  we have plotted entropic densities defined by
  \begin{equation}
     \Psi _{m}^{\gamma}(x) \equiv  -|\psi_m^\gamma(x)|^2
      \ln|\psi_m^\gamma(x)|^2,
  \end{equation}
  and
  \begin{equation}
     \Phi _{m}^{\gamma}(p)\equiv-|\phi_m^\gamma(p)|^2
     \ln|\phi_m^\gamma(p)|^2,
  \end{equation}
  taking into account position and momentum space representation, respectively. The entropy densities  $\Psi _{m}^{\gamma}(x)$ and
  $\Phi _{m}^{\gamma}(p)$ provide a measure of information about localization
  of the particle in the respective spaces.
  The position space entropy densities $\Psi _{m}^{\gamma}(x)$ were depicted graphically versus
  position $x$ in figures 1 and 3 respectively for
  $\gamma=3/2$ and $7/2$. Also, the momentum space entropy densities $\Phi _{m}^{\gamma}(p)$
  are plotted versus $p$ in figures  2 and 4 respectively for
  $\gamma=3/2$ and $7/2$ (notice that the graphs related to the case $\gamma=5/2$, which the numerical results have been displayed in
  Table 1, are omitted due to the economic of space).
  Each group of figures contains different  values of $m$, i.e.,  $1, 2, 3$.
  The general (overall) shape of the figures depend strictly on the
  values of $\gamma$. From the figures 1-a and 3-a, (belong to
  $m=0$), it is clearly seen that in all cases the densities of position
  entropies begin from $0$, then increase with increasing $x$, include some oscillations,
  and finally tend again to zero. Qualitatively, similar situation holds for the set of
  figures "1-b and 3-b" for $m=1$, "1-c and 3-c" for $m=2$, and "1-d and 3-d" for $m=3$.
  Some differences may be mentioned that are listed below: i) the number of
  oscillations grows up with increasing $\gamma$, ii) oscillations
  occurs between $0$ and it's maximum value $\simeq 0.35$ for $m \geq
  1$, with small oscillations occur on the pick of large
  oscillations.

 From the set of figures "2-a and 4-a", have been plotted for
 different values of $\gamma$, but all for $m=0$, it is observed that
 they have maximums at $p=0$ and then tend to $0$ for enough large
 momentum. The set of figures "2-b and 4-b" for $m=1$, "2-c and 4-c" for
 $m=2$, "2-d and 4-d" for $m=3$ show that the densities of the momentum
 entropies have local extremum at $p=0$, in addition to extra
 extremum distributed symmetrically around the vertical axis $p=0$.
 Also,
 the number of oscillations increase with increasing the $m$
 values. However, again the general shapes of the figures do not
 change essentially for each $m$ with different chosen values of $\gamma$. Also, it
 is seen that the maximum heights of the figures of entropic densities, either
 in $x$ or $p$ space are at about $\simeq 0.35 \pm 0.02$.

\vspace{1cm}
 \begin{table}[htp!]
 \caption{\label{math-tab1} Our numerical results which establish the BBM inequality
 for the isotonic oscillator when some fixed values of $m$ and $\gamma$ are considered.}
 \begin{center}
  \begin{tabular}{ccccccccccc }
   \hline
   \hline
   $\gamma$ & $m$ & &$S_{m}^{\gamma}(\rho)$&& & $S_{m}^{\gamma}(\xi)$ & &$S_{m}^{\gamma}
   (\rho)+S_{m}^{\gamma}(\xi)$ & &$1+\ln \pi$ \\
     \hline
 $\frac{3}{2}$& 0 & &0.6496& && 1.5807& & 2.2303 & &2.1447 \\
              & 1 & &0.9166& && 1.9052& & 2.8218 & &2.1447 \\
              & 2 & &1.0749& && 2.0839& & 3.1588 & &2.1447 \\
              & 3 & &1.1889& && 2.2079& & 3.3968 & &2.1447 \\
    \hline
 $\frac{5}{2}$& 0 & &0.6852& && 1.4941& & 2.1793 & &2.1447 \\
              & 1 & &0.9456& && 1.8167& & 2.7623 & &2.1447 \\
              & 2 & &1.0985& && 2.0018& & 3.1003 & &2.1447 \\
              & 3 & &1.2087& && 2.1329& & 3.3416 & &2.1447 \\
    \hline
 $\frac{7}{2}$& 0 & &0.6984& && 1.4663& & 2.1647 & &2.1447 \\
              & 1 & &0.9591& && 1.7797& & 2.7388 & &2.1447 \\
              & 2 & &1.1108& && 1.9628& & 3.0736 & &2.1447 \\
              & 3 & &1.2198&& & 2.2042& & 3.4240 & &2.1447 \\
   \hline
   \hline
 \end{tabular}
\end{center}
\end{table}

\vspace{4cm}
\begin{table}[htp!]
\caption{\label{math-tab2} Our numerical results for calculating
Heisenberg uncertainty relation $(\Delta x)^2(\Delta p)^2\geq
0.25$  for the isotonic oscillator with some fixed values of $m$
and $\gamma$.}
\begin{center}
\begin{tabular}{ccccccccc }
   \hline
   \hline
      $\gamma$ & $m$ && $(\Delta x)^2$& && $(\Delta p)^2$ & $(\Delta x)^2(\Delta p)^2$
       & $min\{(\Delta x)^2(\Delta p)^2\}$ \\
      \hline
 $\frac{3}{2}$& 0& & 0.2268& && 1.4640 & 0.3320 & 0.2500 \\
              & 1 && 0.6352& && 3.4456 & 2.1887 & 0.2500 \\
              & 2& & 1.0238&& & 5.4317 & 5.5608 & 0.2500 \\
              & 3& & 1.4074&& & 7.4199 & 10.4424& 0.2500 \\
    \hline
 $\frac{5}{2}$& 0& & 0.2365&& & 1.1666 & 0.2759 & 0.2500 \\
              & 1& & 0.6746&& & 3.1665 & 2.1362 & 0.2500 \\
              & 2 && 1.0869&& & 5.1663 & 5.6152 & 0.2500 \\
              & 3& & 1.4875&& & 7.1661 & 10.6597& 0.2500 \\
    \hline
 $\frac{7}{2}$& 0 && 0.2405&& & 1.1000 & 0.2646 & 0.2500 \\
              & 1& & 0.6939&& & 3.1000 & 2.1511 & 0.2500 \\
              & 2& & 1.1225&& & 5.1000 & 5.7235 & 0.2500 \\
              & 3 && 1.5367&& & 7.0999& 10.9108& 0.2500 \\
   \hline
   \hline
 \end{tabular}
\end{center}
\end{table}

\section{Conclusion}

 In summary, we investigated squeezing in the position and
momentum entropies, as well as in the quadratures of the
field for the eigenstates of the isotonic oscillator. Note that
'coherent states' corresponding to the isotonic oscillator may
be found in recent literature \cite{{thriasanter}},
in which the authors used
the eigenfunctions of the isotonic oscillator as the basis for
their construction. The coherent states and their even and
odd states associated with the isotonic oscillator have also
been constructed and their non-classical signs were discussed
in \cite{chinees, evenodd}. However, their approaches to the introduction
of non-classical states are essentially different from ours.
Meanwhile, as is well known, the main motivation for the
construction, generalization and generation of various classes
of coherent states lies in searching for non-classical signs in
them. We have explored the non-classical signs, especially
entropic squeezing, in the eigenfunctions of the isotonic
oscillator. The investigation of the information entropy has
been carried out for several solvable non-degenerate quantum
systems \cite{majernik, poshteller, morse},
 but no squeezing effect has been reported.
Summing up, the observation of entropic squeezing, which
occurs for the ground and first excited states (of the isotonic
oscillator), and the quadrature squeezing which occurs for its
ground states, characterizes our contribution from previous
work. Interestingly, the strength of the squeezing for $\psi^\gamma_{m=0} (x)$ tends to a finite value
 $0.25$ with increasing $\gamma$.As a result,
all ground states of the isotonic oscillator are squeezed.
Obviously, these non-classical signs may also be revealed and
observed in the literature for various constructed quantum
states. For instance, we may refer to 'nonlinear coherent
states' which are mostly (quadrature) squeezed \cite{nlcss}, but the relation of the evolved deformation
 function $f(n)$ (which has the central role in this approach) to a "physical potential" is a problem that has
 not been transparent up to know, unless for the special (linear) case of
 $f(n)=1$, which is simply
connected to the harmonic oscillator (where it does not
possess any of the non-classicality features). Even, when
one works with coherent states for the solvable quantum
systems, it is worth noticing that in these cases one does
not deal with the eigenstates of the physical potential,
instead, the states are constructed by some specific definitions
and requirements, and then the non-classical features are
examined \cite{GK}.
 However, we want to emphasize that the
non-classical states are directly related to a special, known physical system. So, the results in this paper are new and
interesting to quantum opticians. As a final point, the constant
$A$ in the Hamiltonian (5) characterizes the relative strength of the
 $1/r^2$, i.e. centripetal potential. In a rotating diatomic, for
example, this constant would be determined approximately by
the rotational energy level. It may be interesting to consider
the connection between squeezing in the information entropy
and the rotational constant $A$, to motivate future experimental
studies of these results.


\section*{Acknowledgement}
    One of the authors (MKT) would like to thank Professor
    R Roknizadeh from the Quantum Optics Group of The
    University of Isfahan for useful discussions.

\vspace{2cm}


 \vspace{1.cm}

 {\bf Figure Captions:}
 \vspace{.5cm}

 {\bf Fig. 1}  {The position space entropy densities of the isotonic oscillator for m=0, 1, 2, 3 and $\gamma =3/2$.}

 \vspace{.2cm}

   {\bf Fig. 2} {The momentum space entropy densities of the isotonic oscillator for m=0, 1, 2, 3 and $\gamma =3/2$.}

 \vspace{.2cm}

   {\bf Fig. 3} {The position space entropy densities of the isotonic oscillator for m=0, 1, 2, 3 and
   $\gamma =7/2$.}

 \vspace{.2cm}

  {\bf Fig. 4} {The momentum space entropy densities of the isotonic oscillator for m=0, 1, 2, 3 and $\gamma =7/2$.}

  \end{document}